\documentclass[3p,times,procedia]{elsarticle}

\biboptions{sort&compress}

\usepackage{amssymb}

\usepackage[figuresright]{rotating}

\newcommand{\ee}{$e^{+}e^{-}$}

\newcommand{\detadphi}{$\Delta\eta\Delta\varphi$}

\begin{document}

\begin{frontmatter}




\title{Influence of quantum conservation laws on particle production in hadron collisions}

\author[label1]{Ma\l{}gorzata Anna Janik}
\author[label1]{\L{}ukasz Kamil Graczykowski}
\author[label1]{Adam Kisiel}
\address[label1]{Warsaw University of Technology, Faculty of Physics, Koszykowa 75, PL-00-662 Warszawa, Poland}

\author{}

\address{}

\begin{abstract}
Conservation laws strongly influence production of particles in high-energy particle collisions. Effects connected to these mechanisms were studied in details using correlation techniques in \ee\ collisions. At the time, models were tuned to correctly reproduce the measurements. Similar studies for hadron-hadron collisions have never been performed, until recent ALICE measurements. ALICE has reported on studies of untriggered two-particle angular correlations of identified particles ($\pi$, $K$, and p) measured in pp collisions at center-of-mass energy of $\sqrt{s}=$ 7 TeV. Those preliminary results confirm that also in hadron-hadron collisions, at much higher energies, conservation laws strongly influence the shape of the correlation functions for different particle types and must be taken into account while analysing the data. Moreover, they show that the contemporary models (PYTHIA, PHOJET) no longer reproduce the experimental data well.

\end{abstract}


\end{frontmatter}


\section{Introduction}
\label{sec:intro}
One of the most important ingredients of the theoretical description of relativistic particle collisions is the particle production mechanism. Currently, there are few models describing such processes (e.g. string fragmentation model \cite{Andersson:1998tv}, also referred to as the "Lund model", incorporated in PYTHIA Monte Carlo generator \cite{Sjostrand:2006za}). However, the implementation details of such theoretical calculations depend strongly on the experimental data which are used to constrain their input parameters, especially in the non-perturbative regime at low transverse momenta of the produced particles. For example, some of the parameters in the current version of PYTHIA, particularly those describing production of baryons, have not been improved for many years (since \ee\ collisions at much lower energies). 
This fact is reflected in the comparison of the model predictions with the yet preliminary experimental results of angular correlations in 
\detadphi\ space (relative pseudorapidity and azimuthal angle)  with the results from model predictions. Current models explain correlations of mesons only qualitatively, but in the case of baryons they fail to describe experimental data both qualitatively and quantitatively; anti-correlation is observed in experimental results in contrast to a correlation observed in models. This newly discovered scientific problem has fundamental impact for the theoretical description of particle production in high-energy physics. 

\vspace{-0.3cm}
\section{Two-particle rapidity correlations in $e^{+}e^{-}$ collisions}
\label{sec:epluseminus}
Studies of the particle production mechanism in elementary collisions date back to the times of R. Feynman and R. Field, who proposed a simple mechanism describing the principles of the creation of the so-called “jets” (collimated streams of particles) in 1977 \cite{Field:1977fa, Field:2015hpa}. They proposed rules on how the particles are produced, how the energy is distributed and considered limitations connected with the conservation laws.

Elements of the proposed scheme are used even today in the most popular fragmentation models. However, the implementation details have to be connected with the experimental data. It is then a task for the experiment to provide basic information: how strong the correlations between the created hadrons should be? How does this correlation change, when we create two or more baryons or strange particles in a single parton fragmentation? Answers to these and other questions have been searched, so far, only in \ee\ collisions, of much lower energies and on substantially smaller data samples  \cite{Aihara:1986fy, Althoff:1984ut, Abreu:1997mp, Acton:1993ux}. 

In Ref.~\cite{Aihara:1986fy}, two-particle rapidity correlations are reported for \ee\ collisions at 29 GeV. It was observed that pairs of particles with opposite baryon number (p-$\mathrm{\bar{p}}$, p-$\mathrm{\bar{\Lambda}}$, $\Lambda$-$\mathrm{\bar{\Lambda}}$) produce significant correlation while pairs of particles with the same baryon number ($\mathrm{\bar{p}}$-$\mathrm{\bar{p}}$, $\mathrm{\bar{p}}$-$\mathrm{\bar{\Lambda}}$, $\mathrm{\bar{\Lambda}}$-$\mathrm{\bar{\Lambda}}$) produce significant anti-correlations. Moreover, in the same article it was reported that mechanisms ensuring that the baryon number is conserved not only globally, for the whole event, but also locally for each parton fragmentation are crucial for reproducing the experimentally observed anti-correlation. Calculations employing the Lund model incorporating local compensation of baryon number described \ee\ data (both correlation and anti-correlation) very well. 

\vspace{-0.3cm}
\section{Two-particle $\Delta\eta\Delta\varphi$ correlations in pp collisions}
\label{sec:detadphi}
\vspace{-0.1cm}
ALICE has reported the results of the angular correlation measurements from pp collisions in Ref.~\cite{Graczykowski:2014eqa,Janik:2014cua,Janik:2012ya}. In this article we would like to focus on the analysis of identified particles ($\pi$, $K$, p); see ALICE preliminary results from Ref.~\cite{QM2015slides}. These results show that the magnitude of the near-side peak (positive correlation structure centered at $(\Delta\eta,\Delta\varphi) = (0,0)$) change between different particle species. The observed qualitative differences can be explained with conservation laws applied locally for each single fragmentation: energy, momentum, charge, strangeness, and baryon number. 

\begin{figure}[ht!]
       	\includegraphics[width=0.95\textwidth]{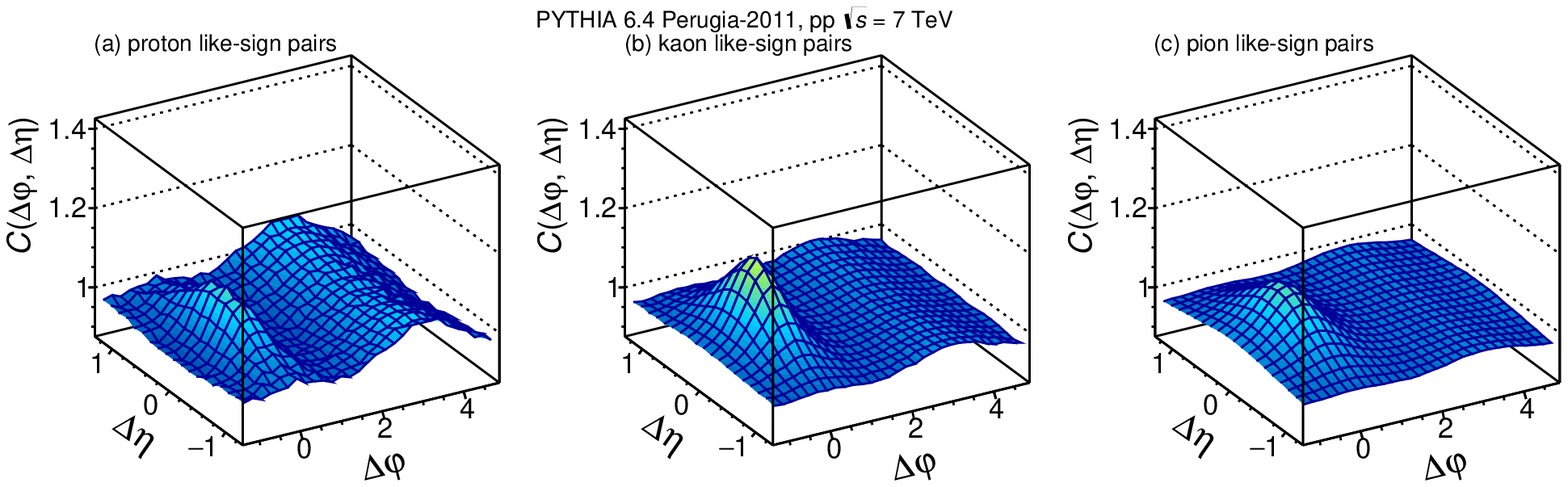}
       	\includegraphics[width=0.95\textwidth]{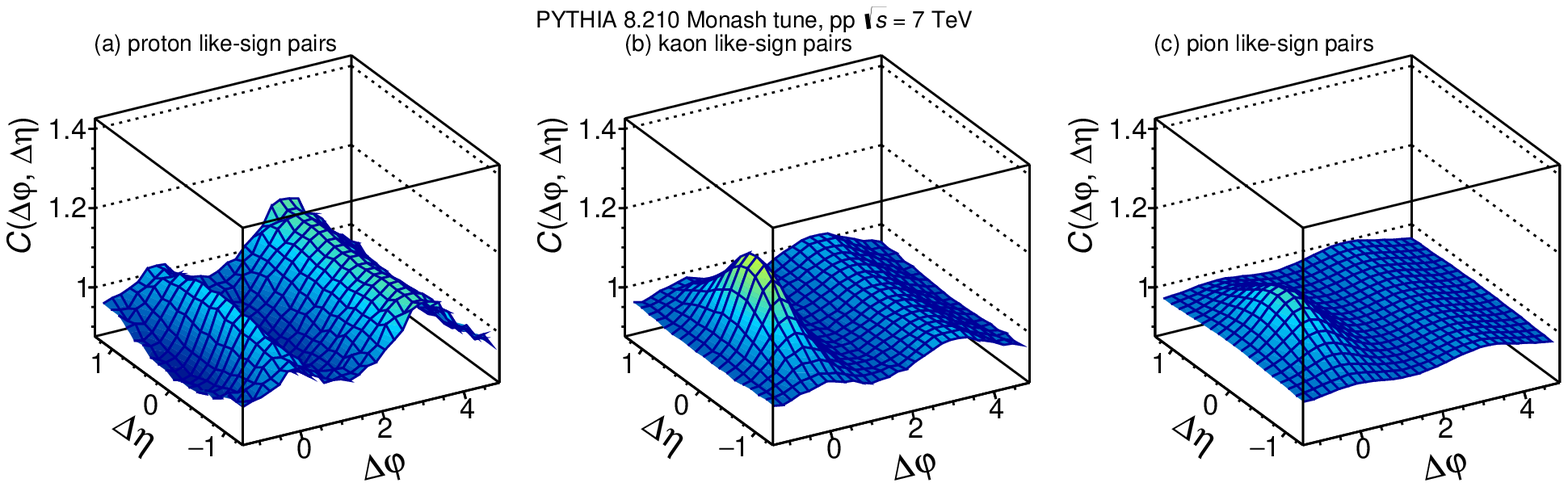}
       	\includegraphics[width=0.95\textwidth]{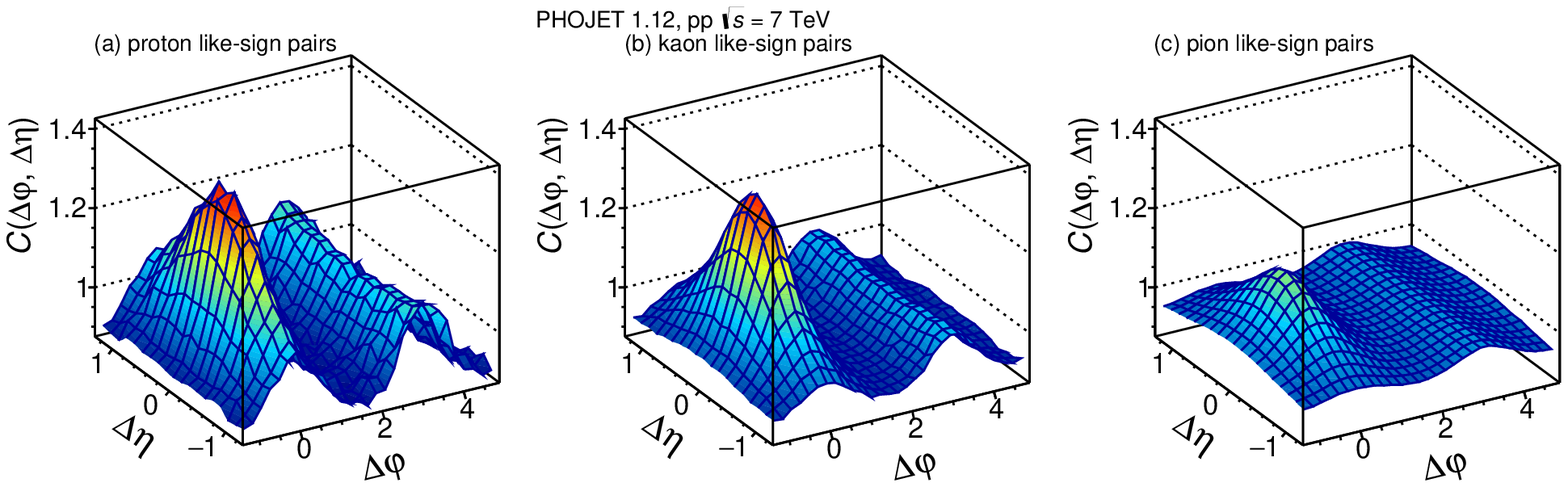}
       	\vspace{-0.2cm}
       	\caption{ Theoretical correlation functions for like-sign pairs of (a) protons, (b) kaons, (c) pions for pp collisions at 7 TeV from top: PYTHIA Perugia-2011, middle: PYTHIA 8 Monash tune, bottom: PHOJET 1.12 generators; they all show the near-side peak for like-sign protons, in contrast to the data, which shown an anti-correlation; see Ref.~\cite{QM2015slides}. }
       	\label{Fig:PYTHIA}
       \end{figure}

Most importantly, it should be noted that ALICE results also show an anti-correlation for like-sign protons, similarly as in  \ee\ collisions. This finding is surprising, since at LHC energies we expect jet production to strongly suppress anti-correlation effects (production of correlated protons should be much more abundant). Such expectations were also supported by the models (see Fig.~\ref{Fig:PYTHIA} for three examples): it can be seen that in all theoretical simulations significant correlation, the near-side peak, is seen for like-sign protons. Similar discrepancy can be seen for correlation functions of unlike-sign protons; 
the magnitude of the near-side peak in theoretical models is nearly twice as large as in experimental data.
This comparison suggests that in models significantly more correlated protons are being produced than in the experiment. 

       \begin{figure}[ht!]
       	\includegraphics[width=0.95\textwidth]{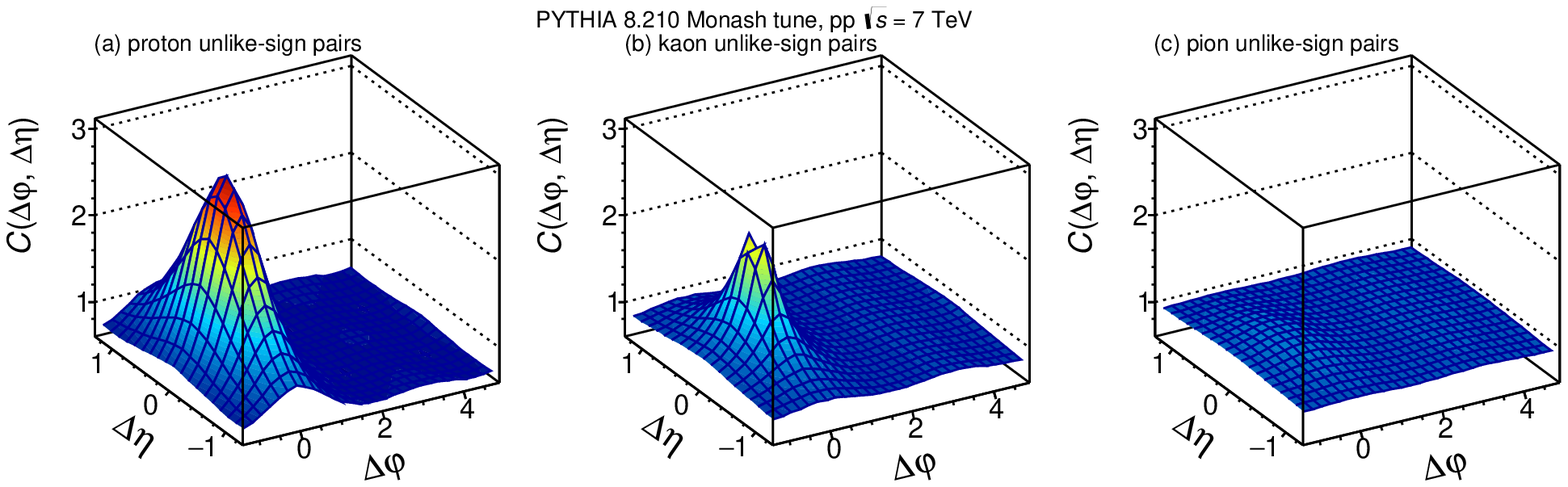}
       	\vspace{-0.2cm}
       	\caption{ Theoretical correlation functions for unlike-sign pairs of (a) protons, (b) kaons, (c) pions for pp collisions at 7 TeV from PYTHIA 8 Monash tune.  Models suggest that significantly more correlated protons are being produced than it is suggested by data; compare with Ref.~\cite{QM2015slides}. }
	   	\label{Fig:PYTHIAunlike}
       \end{figure}
       
In summary, in contrast to  \ee\ collisions at $\sqrt{s}=29$~GeV, at LHC energies the two-particle correlations measured by ALICE in pp collisions are not described by the Monte Carlo models. Moreover, these results suggest that production of baryon-baryon (or antibaryon-antibaryon) pairs in a single jet/minijet is suppressed.  






\vspace{-0.3cm}
\section{ConservAtion Laws Model}
\label{sec:calm}
\begin{figure}
    \centering
    	\includegraphics[width=0.55\textwidth]{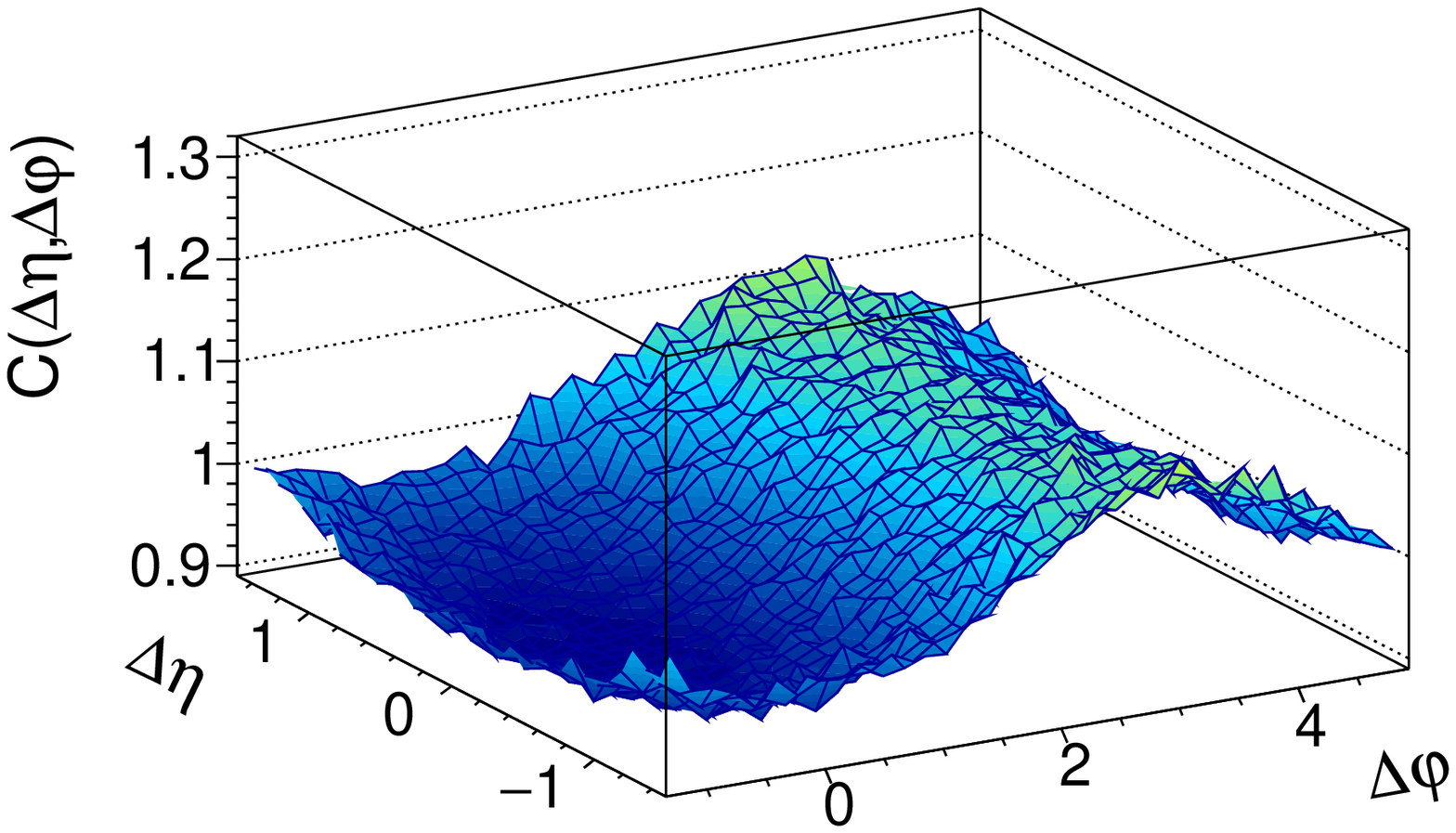}
    	\caption{Correlation function of charged particles obtained from CALM model including \textbf{only} conservation laws.}
    	\label{Fig:genbod_pions0only}
    \end{figure}

\vspace{-0.1cm}
As a basic step towards understanding the role of conservation laws and their influence on the analyzed observables, we studied events in which energy and momentum are conserved and no other physics mechanisms are involved. For such a case a simple Monte Carlo model was developed in order to explore the impact of the conservation laws on the correlation functions -- CALM (ConservAtion Laws Model). CALM allows us to generate events in which only energy, momentum and quantum numbers local to the emission from boosted clusters are conserved and no other processes are involved\footnote{ N-body Monte Carlo event generators are used for computing the phase space distribution of particles produced in the collision \cite{GENBOD68,Meres:2011ek}}. The model also reproduces the usual jet/minijet correlation shape with the near-side peak and the away-side ridge. 
In Fig.~\ref{Fig:genbod_pions0only} the correlation function presenting neutral pions distributed isotropically in the whole phase-space, with momentum conservation as the only constraint, is shown. Such a simple description reproduces qualitatively structures observed for like-sign protons in ALICE data presented in Ref.~\cite{QM2015slides}.

\vspace{-0.3cm}
\section{Summary}
\label{sec:summary}
\vspace{-0.1cm}
We expect that as a result of the analysis of two-particle angular correlations in ALICE novel information on particle production, both mesons ($\pi$, $K$) and baryons (p), can be obtained. Preliminary results of the experimental data analysis (in pp collisions) exhibit significant qualitative differences with respect to theoretical predictions calculated from models -- they are especially visible for baryons. We are able to reproduce the interesting structure observed in ALICE with a new model, CALM. Further analysis of the results may be of a key significance for the development of the theoretical description of particle collisions and may trigger a need for major modifications of particle production models.

\vspace{-0.3cm}
\section*{Acknowledgements}
\label{sec:acknowledgements}
\vspace{-0.1cm}
This research has been financed by the Polish National Science Centre in Poland, based on the decisions no. 
DEC-2014/13/B/ST2/04054 and DEC-2013/08/M/ST2/00598.


    






\vspace{-0.2cm}
\bibliographystyle{elsarticle-num}
\bibliography{bibliography}

\begin{thebibliography}{10}
\expandafter\ifx\csname url\endcsname\relax
  \def\url#1{\texttt{#1}}\fi
\expandafter\ifx\csname urlprefix\endcsname\relax\def\urlprefix{URL }\fi
\expandafter\ifx\csname href\endcsname\relax
  \def\href#1#2{#2} \def\path#1{#1}\fi

\bibitem{Andersson:1998tv}
B.~Andersson, {The Lund model}, Camb.Monogr.Part.Phys.Nucl.Phys.Cosmol. 7
  (1997) 1--471.

\bibitem{Sjostrand:2006za}
T.~Sjöstrand, S.~Mrenna, P.~Z. Skands, {PYTHIA 6.4 Physics and Manual}, JHEP
  0605 (2006) 026.
\newblock \href {http://arxiv.org/abs/hep-ph/0603175}
  {\path{arXiv:hep-ph/0603175}}.

\bibitem{Field:1977fa}
R.~Field, R.~Feynman, {A Parametrization of the Properties of Quark Jets},
  Nucl.Phys. B136 (1978) 1.

\bibitem{Field:2015hpa}
R.~Field, {Quark elastic scattering as a source of high transverse momentum
  mesons}, Int. J. Mod. Phys. A30~(01) (2015) 1530005.

\bibitem{Aihara:1986fy}
H.~Aihara, et~al., {Study of baryon correlations in $e^+e^-$ annihilation at 29
  GeV}, Phys.Rev.Lett. 57 (1986) 3140.

\bibitem{Althoff:1984ut}
M.~Althoff, et~al., {Evidence for Local Compensation of Baryon Number in $e^+
  e^-$ Annihilation}, Phys.Lett. B139 (1984) 126.

\bibitem{Abreu:1997mp}
P.~Abreu, et~al., {Rapidity correlations in Lambda baryon and proton production
  in hadronic Z0 decays}, Phys.Lett. B416 (1998) 247--256.
\newblock \href {http://dx.doi.org/10.1016/S0370-2693(97)01380-4}
  {\path{doi:10.1016/S0370-2693(97)01380-4}}.

\bibitem{Acton:1993ux}
P.~Acton, et~al., {Evidence for chain-like production of strange baryon pairs
  in jets}, Phys.Lett. B305 (1993) 415--427.

\bibitem{Graczykowski:2014eqa}
{\L}.~K. Graczykowski, M.~A. Janik, {Angular correlations measured in pp
  collisions by ALICE at the LHC}, Nucl.Phys. A926 (2014) 205--212.
\newblock \href {http://arxiv.org/abs/1401.4306} {\path{arXiv:1401.4306}},
  \href {http://dx.doi.org/10.1016/j.nuclphysa.2014.03.004}
  {\path{doi:10.1016/j.nuclphysa.2014.03.004}}.

\bibitem{Janik:2014cua}
M.~A. Janik, {Two-particle angular correlations in pp collisions recorded with
  the ALICE detector at the LHC}, EPJ Web Conf. 71 (2014) 00058.
\newblock \href {http://arxiv.org/abs/1402.3988} {\path{arXiv:1402.3988}}.

\bibitem{Janik:2012ya}
M.~A. Janik, {$\Delta\eta\Delta\phi$ angular correlations in pp collisions at
  the LHC registered by the ALICE experiment}, PoS WPCF2011 (2011) 026.
\newblock \href {http://arxiv.org/abs/1203.2844} {\path{arXiv:1203.2844}}.

\bibitem{QM2015slides}
M.~Janik, L.~Graczykowski, A.~Kisiel,
  \href{https://indico.cern.ch/event/355454/session/59/contribution/753}{{Influence
  of quantum conservation laws on particle production in hadron collisions}}
  (2015).
\newblock \href {http://arxiv.org/abs/QM2015, Kobe, slides}
  {\path{QM2015, Kobe, slides}}.
\newline\urlprefix\url{https://indico.cern.ch/event/355454/session/59/contribution/753}

\bibitem{GENBOD68}
F.~E. James, {Monte Carlo phase space}, CERN, Geneva, 1968, p.~41.

\bibitem{Meres:2011ek}
M.~Meres, I.~Melo, B.~Tomasik, V.~Balek, V.~Cerny, {Generating heavy particles
  with energy and momentum conservation}, Comput. Phys. Commun. 182 (2011)
  2561--2566.
\newblock \href {http://arxiv.org/abs/1101.3339} {\path{arXiv:1101.3339}},
  \href {http://dx.doi.org/10.1016/j.cpc.2011.06.015}
  {\path{doi:10.1016/j.cpc.2011.06.015}}.

\end{thebibliography}







\end{document}